\documentclass[onecolumn,showpacs,superscriptaddress,showkeys,preprint,prb]{revtex4-1}

\usepackage{amssymb, amsmath}
\usepackage[english]{babel}
\usepackage{bm}
\usepackage{graphicx}
\usepackage{lmodern}
\usepackage[utf8]{inputenc}
\usepackage{color}
\usepackage[dvipsnames]{xcolor}
\usepackage{mciteplus}
\usepackage{comment}

\begin{document}


\title{Superradiant to subradiant phase transition in the open system Dicke model: Dark state cascades}
\author{Michael Gegg}
\email[]{michael.gegg@tu-berlin.de}
\affiliation{Institut für Theoretische Physik, Nichtlineare Optik und
Quantenelektronik, Technische Universität Berlin, Hardenbergstr. 36, EW 7-1, 10623
Berlin, Germany}
\author{Alexander Carmele}
\affiliation{Institut für Theoretische Physik, Nichtlineare Optik und
Quantenelektronik, Technische Universität Berlin, Hardenbergstr. 36, EW 7-1, 10623
Berlin, Germany}
\author{Andreas Knorr}
\affiliation{Institut für Theoretische Physik, Nichtlineare Optik und
Quantenelektronik, Technische Universität Berlin, Hardenbergstr. 36, EW 7-1, 10623
Berlin, Germany}
\author{Marten Richter}
\affiliation{Institut für Theoretische Physik, Nichtlineare Optik und
Quantenelektronik, Technische Universität Berlin, Hardenbergstr. 36, EW 7-1, 10623
Berlin, Germany}

\begin{abstract}
Collectivity in ensembles of atoms gives rise to effects like super- and subradiance. While superradiance is well studied and experimentally accessible, subradiance remains elusive since it is difficult to track experimentally as well as theoretically. Here we present a new type of phase transition in the resonantly driven, open Dicke model that leads to a deterministic generation of subradiant states. At the transition the system switches from a predominantly superradiant to a predominantly subradiant state. Clear experimental signatures for the effect are presented and entanglement properties are discussed. Letting the system relax into the ground state generates a cascade of dark Dicke states, with dark state populations up to unity. Furthermore we introduce a collectivity measure that allows to quantify collective behavior.
 \end{abstract}


\date{\today}

\maketitle
 

\section{Introduction}

The open (and closed) system Dicke model has been a work horse in quantum optics and beyond for decades \cite{Dicke:PhysRev:54,Garraway:PTA:11,Wang:PhysRevA:73,Walls:SPTP:78,Carmichael:JPhysB:80,Drummond:PhysRevA:81,Hassan:PhysA:80,Lawande:JPhysB:81,Emary:PhysRevLett:03,Schneider:PhysRevA:02,Schneebeli:PhysRevLett:08,Lee:PhysRevLett:12,Gonzales:PhysRevLett:13,Su:PRL:13,Carr:PhysRevLett:13,Genway:PhysRevLett:14,Zou:PhysRevLett:14,Richter:PhysRevB:15,Wolfe:PhysRevLett:14,Scully:PhysRevLett:15,Cong:JOSAB:16,Kirton:PhysRevLett:17}. Current research on Dicke model based systems includes novel laser-like systems \cite{Richter:PhysRevB:15}, phase transitions \cite{Carr:PhysRevLett:13,Kirton:PhysRevLett:17}, quantum information and super/subradiance \cite{Schneider:PhysRevA:02,Gonzales:PhysRevLett:13,Wolfe:PhysRevLett:14,Scully:PhysRevLett:15,Guerin:PhysRevLett:16}. In recent years superradiance has been investigated with respect to entanglement \cite{Wolfe:PhysRevLett:14} and subradiance for its prospects to store quantum information \cite{Scully:PhysRevLett:15,Guerin:PhysRevLett:16}. The Dicke model assumes $N$ identical two-level systems, interacting with a bosonic cavity mode.\\
Investigating subradiant effects in a consistent open system theory was not feasible for a long time since in a straight forward approach the master equation scales exponentially in the number $N$ of two-level systems.  This renders full simulations even for small $N$ impossible, however subradiance is a few and many particle effect. Existing limits and approximations for both analytical and numerical treatments addressing this problem are not suited to study subradiance for moderate $N$ \cite{Walls:SPTP:78,Carmichael:JPhysB:80,Drummond:PhysRevA:81,Hassan:PhysA:80,Lawande:JPhysB:81,Schneider:PhysRevA:02,Gonzales:PhysRevLett:13}. Usually for superradiance total spin conservation (explained below) is assumed, entirely neglecting subradiant states. This reduces the numerical complexity to $\sim N^2$ or sometimes even allows analytic solutions \cite{Gonzales:PhysRevLett:13,Hassan:PhysA:80,Lawande:JPhysB:81}. However ubiquitous phenomena in real systems like decay processes and pure dephasing break this symmetry. Therefore, both realistic treatments and subradiant effects require a different methodology.\\
Symmetries in the associated master equations reduce the complexity from an exponential scaling in $N$ to a polynomial scaling $\sim N^3$, even without total spin conservation \cite{Richter:PhysRevB:15,Gegg:NJP:16,Hartmann:QIC:16,Xu:PhysRevA:13,Kirton:PhysRevLett:17}. This makes exact calculations for moderate emitter numbers feasible and removes constraints imposed by assumptions and approximations. Furthermore the method can be applied to any permutation symmetric multi-level system setup \cite{Gegg:NJP:16}.\\
In this work we investigate the population of subradiant states through decay and pure dephasing processes -- both do not conserve the total spin. Counterintuitively, the cavity lifetime determines the population of the subradiant states: Increasing the external driving results in a nonequilibrium phase transition and for short cavity lifetimes subradiant states are always suppressed by quantum coherence. Contrary increasing the cavity lifetime results in an amplification of subradiant states due to quantum coherence. Experimentally accessible signatures of this effect and entanglement properties via spin squeezing are discussed. Switching off the external driving, the subsequent relaxation into the ground state forms a long-lived cascade of dark Dicke states. This results in a simple, deterministic protocol for dark state preparation with populations close to unity under the influence of dephasing, with applications in quantum information storage.\\
\begin{figure}
\centering
 \includegraphics[scale=1.4]{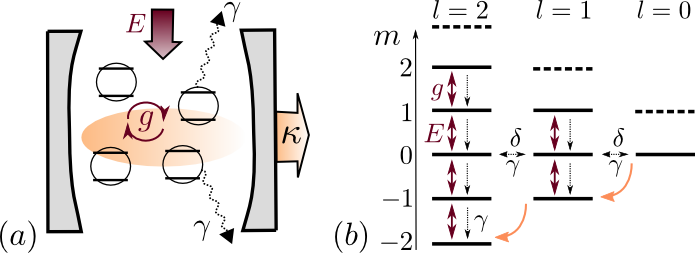}
 \caption{Illustrating the open Dicke model: (a) Schematic representation of the system. (b) Dicke states for $N=4$. The lowest state in each $l$ subspace is dark -- the lowest state in the superradiant $l_{max} = N/2$ subspace is the ground state. The interactions are depicted: Hamiltonian part (purple,thick), dissipators $\mathcal{D}_{de}$ and $\mathcal{D}_{pd}$ (black,thin) and dark state cascade (orange,curved). Dashed lines indicate the additional states for $N=5$ (with different values of $m,l$).}
 \label{fig.systemsketch}
\end{figure}

\section{Model system}
We consider the usual Dicke model with an additional classical optical, cw field $E$ driving all TLS identically. Driving is necessary since subradiant states are excited states. In a frame rotating at the external laser frequency, using the rotating wave approximation the system Hamiltonian reads
\begin{equation}
H = \hbar\Delta_0 b^\dagger b+ \hbar\Delta_1 J_{11} + \hbar g (J_{10} b + J_{01} b^\dagger) + \hbar E (J_{10} + J_{01}),
\label{eq.ham}
\end{equation}
where $\Delta_0$, $\Delta_1$ are the mode and TLS detuning, $g$ is the TLS-mode coupling, $E$ is the optical driving, $b,b^\dagger$ are photonic operators and $J_k = \sum_i \sigma_{k}^i$, $k = 11,10,01,00$ are the collective spin operators. Excited and ground state of the individual TLS $i$ are $|1\rangle_i$, $|0\rangle_i$, the lower indices of the spin operators represent the Ket and Bra notation: $\sigma_{11}^i = |1\rangle_i \langle 1|_i$, $\sigma_{10}^i = |1\rangle_i \langle 0|_i$, $\sigma_{01}^i = |0\rangle_i \langle 1|_i$ and $\sigma_{00}^i = |0\rangle_i \langle 0|_i$. We assume resonant excitation field, cavity and TLS.
Both cavity and TLS are subject to loss and dephasing, using Lindblad formalism \cite{Breuer::02}. The master equation reads
\begin{equation}
\partial_t \rho = \mathcal{L}\rho = \frac{i}{\hbar}[\rho,H] + \mathcal{D}_{de}(\rho) + \mathcal{D}_{pd}(\rho) + \mathcal{D}_{ph}(\rho).
\label{eq.qme}
\end{equation}
The Lindblad dissipators describe decay processes like individual radiative and non-radiative decay $\mathcal{D}_{de}(\rho) = \gamma/2\sum_i (2 \sigma_{01}^i \rho \sigma_{10}^i - \sigma_{11}^i \rho - \rho \sigma_{11}^i)$, pure dephasing $\mathcal{D}_{pd}(\rho) = \delta/2 \sum_i ( \sigma_{z}^i \rho \sigma_{z}^i - \rho )$ and cavity decay $\mathcal{D}_{ph}(\rho)  = \kappa/2(2 b \rho b^\dagger - b^\dagger b \rho - \rho b^\dagger b)$, see Fig. \ref{fig.systemsketch} (a). We use $\sigma_{z}^i = \sigma_{11}^i -\sigma_{00}^i$.  All contributions to the master equation except $\mathcal{D}_{de}$ and $\mathcal{D}_{pd}$ are total spin preserving, (Fig. \ref{fig.systemsketch} (b)).
The total spin $l(l+1)$ is the eigenvalue of the $J^2 = (J_{10}J_{01} + J_{01}J_{10})/2 + J_z$ operator, with $J_z = 1/2\sum_i \sigma_{z}^i$. The value of $l$ varies between $l_{max}=N/2$ for the superradiant subspace and $l_{min}=0,1/2$ for the (most) subradiant subspace. The $J^2$ and $J_z$ eigenvalues determine the coupling strength of the multi TLS (Dicke) state to an optical mode, the collective dipole transition element. This coupling strength distinguishes between superradiance and subradiance. For superradiant states the dipole element scales superlinear in $N$, while for subradiant states the scaling is sublinear in $N$ and some subradiant states are dark \cite{Mandel::95}. Dark means that the dipole transition element of the collective excitation vanishes, meaning these states cannot decay e.g. creating a cavity photon. However these states still decay into other states via the decay and dephasing processes $\mathcal{D}_{de}$ and $\mathcal{D}_{pd}$ acting individually on the emitters, c.f. Fig. \ref{fig.systemsketch} (b). Generally the spin preserving contributions in the master equation (like Eq. \eqref{eq.ham}) generate quantum correlations leading to collective TLS behavior (\emph{both} super- and subradiance are collective effects) and the nonpreserving terms destroy correlations leading to individualization (all properties scale exactly linear in $N$). However only the spin nonpreserving contributions introduce coupling between superradiant and subradiant states, thus in order to prepare subradiant states an interplay of collectivity and individualization is necessary.\\
In the bad cavity limit ($\kappa \gg g$) equation \eqref{eq.qme} corresponds to the cooperative resonance fluorescence setup \cite{Walls:SPTP:78,Carmichael:JPhysB:80}. The system exhibits a non-equilibrium phase transition for increasing $E$ for both total spin preserving and nonpreserving setups, where the nonpreserving setup was studied using mean field theory \cite{Walls:SPTP:78}. For longer cavity lifetimes $\kappa$ the system more and more resembles the absorptive optical bistability setup \cite{Gibbs::85} (instead of driving the TLS, in optical bistability the cavity is driven, opposed to Fig. \ref{fig.systemsketch} (a)). In the range investigated in this work ($\kappa\sim g$) the clear distinction between cooperative resonance fluorescence and optical bistability breaks down, thus combining these distinct fields of quantum optics. Besides the steady state, density matrix states with very long lifetimes can exist in these systems, which lead to the observation of bistabilities in experiments with finite measurement time \cite{Rodriguez:arxiv:16}. In some limits these lifetimes go to infinity, resulting in a second steady state. For optical bistability these long lifetimes are called tunneling times \cite{Sarkar:EuroPhysLett:87,Schenzle:OptComm:79}, more general this phenomenon is called dissipative phase transition \cite{Kessler:PhysRevA:12}.\\

\section{Permutation symmetric method} 
The permutation symmetry allows the incorporation of the individual TLS decay and dephasing while having moderate numbers of TLS and photonic Fock states. Furthermore the states introduced in this method allow a more intuitive understanding of the processes in the system \cite{Hartmann:QIC:16,Gegg:NJP:16}. For a permutation symmetric master equation the TLS part of the density matrix is described by elements $\mathcal{P}[n,k,l]$ with $0 \leq n+k+l \leq N$. These elements describe the full density matrix and their number scales with $\propto N^3$.
For element $\mathcal{P}[n,k,l]$ $n$ of the $N$ TLS are in a $\sigma_{11}$ Liouville state, $k$ are in a $\sigma_{10}$ state and $l$ in a $\sigma_{01}$ state. $\sigma_{01}$ and $\sigma_{10}$ ($k\neq 0$ and/or $l\neq 0$) correspond to a quantum coherence/offdiagonal element in the density matrix.
The different elements can be interpreted as follows: $\mathcal{P}[n,0,0]$ is the incoherent probability of finding the $N$ TLS system with $n$ excited TLS. For instance preparing the system in a thermal state results in a thermal distribution in the $\mathcal{P}[n,0,0]$ densities, or preparing the system in the ground state is equivalent to $\mathcal{P}[0,0,0] = 1$ and zero for all other $\mathcal{P}[n,k,l]$. The elements $\mathcal{P}[n,k,l]$ for $k,l\neq 0$ describe quantum correlations and thus are constructed from the offdiagonal elements of the density matrix. For $k=l$ i.e. $\mathcal{P}[n,k,k]$ these elements are real valued but still represent offdiagonal density matrix elements/coherences in the basis of individual TLS. The $\mathcal{P}[n,k,k]$ are collective quantum contributions and contribute to the \emph{collective Dicke state population} of excited states in the system: More precise the $\mathcal{P}[n,k,k]$ distinguish collective Dicke state populations from classical, individual excited state populations in the open system, density matrix setting: The offdiagonal elements of the density matrix in the individual TLS basis are directly connected to the collective effects in the many emitter setup. In the following we will explain this relation in more detail and introduce a measure to distinguish collective Dicke behavior from classical, individual behavior in the presence of dephasing. For more details on the permutation symmetric method and the density matrix elements $\mathcal{P}[n,k,l]$ please refer to \ref{method} and Refs. \cite{Gegg:NJP:16,psiquasp,github}.\\ 

\section{Collectivity measure} 
Investigating super- and subradiant states requires a suitable measure. Unfortunately computing the respective Dicke state populations is not sufficient for investigating collective effects and quantum coherence, if dephasing is present: Dicke states $|l,m\rangle$ are eigenstates of $J^2$ and $J_z$ with corresponding quantum numbers $l(l+1)$, $0\leq l\leq N/2$ and $|m|\leq l$. $l_{max}=N/2$ defines the superradiant subspace and $l_{min} = 0,1/2$ defines the (most) subradiant subspace, see Fig. \ref{fig.systemsketch} (b). As an example consider the $N=2$ Dicke (or Bell) states: The superradiant subspace consists of three states $|1,-1\rangle$, $|1,0\rangle$, $|1,+1\rangle$ while the subradiant subspace consists of a single dark state $|0,0\rangle$. First we calculate the population of the Dicke states: $\mbox{tr}[|l,m\rangle \langle l,m|\rho]= \big\langle |l,m\rangle \langle l,m| \big\rangle = p(l,m)$ in the local basis.
Using the permutation symmetric density matrix elements $\mathcal{P}[n,k,l]$ we can write these populations as
\begin{equation}
p(l,m) = a_0(l,m) \mathcal{P}[n,0,0] \pm a_1(l,m) \mathcal{P}[n-1,1,1] \ldots,
\end{equation}
with $n = m + N/2$. In the presence of dephasing the elements $\mathcal{P}[n,k,k]$ for $k \neq 0$ (representing quantum coherences) experience dephasing. If the dephasing is strong enough it will completely suppress quantum correlations, i.e. $\mathcal{P}[n,k,k] = 0$ for $k\neq 0$. This represents a completely incoherent mixture of TLS occupations.
For varying numbers of TLS, $\mathcal{P}[n,0,0]$ distributions allow a large variety of populations in super- and subradiant states even if quantum coherences are absent. Generally -- when spin non-conserving terms are included -- the superradiant subspace population decreases, since for large $N$ the superradiant subspace is very small compared to the full Hilbert space ($N+1$ vs. $2^N$). However without quantum coherences ($\mathcal{P}[n,k,k],~k\neq 0$) in the TLS basis the label super- and subradiance becomes meaningless, since the quantum coherences in the local basis are the signatures of the collectivity of the Dicke states and reflect the redistribution of oscillator strength through collective effects (phase locking). Thus -- in the open Dicke model -- $\mathcal{P}[n-k,k,k]$ are the key quantities that distinguish a super- or subradiant state from a classical, incoherent mixture of TLS population ($\mathcal{P}[i,j,k]=0$ for $j,k \neq 0$). The decay process $\mathcal{D}_{de}$ and the pure dephasing $\mathcal{D}_{pd}$ act individually on every TLS and thus destroy the collectivity, resulting in incoherent mixtures.\\
To quantify the effect of collectivity and distinguish between collective (super- and subradiance) and individual (dipole moment scales linear in $N$) behavior we introduce the ratio between the full Dicke subspace population and its incoherent part 
\begin{equation}
R(l) = \frac{\sum_m p(l,m)}{\sum_m a_0(l,m) \mathcal{P}[m+N/2,0,0]},
\end{equation}
as a collectivity measure for the different Dicke subspaces $l$. $R(l)=1$ holds if the influence of quantum correlations between the individual TLS on the subspace population is zero or negligible -- the TLS act \emph{individually}. $R(l)<1$/$R(l)>1$ holds if quantum correlations \emph{suppress}/\emph{increase} the respective subspace occupation -- the TLS act \emph{collectively}. $R(l)$ provides a reality check, since in any experiment dephasing is present and isolated Dicke subspaces (or states) never occur.\\
\begin{figure}[t]
\centering
\includegraphics[scale=0.8]{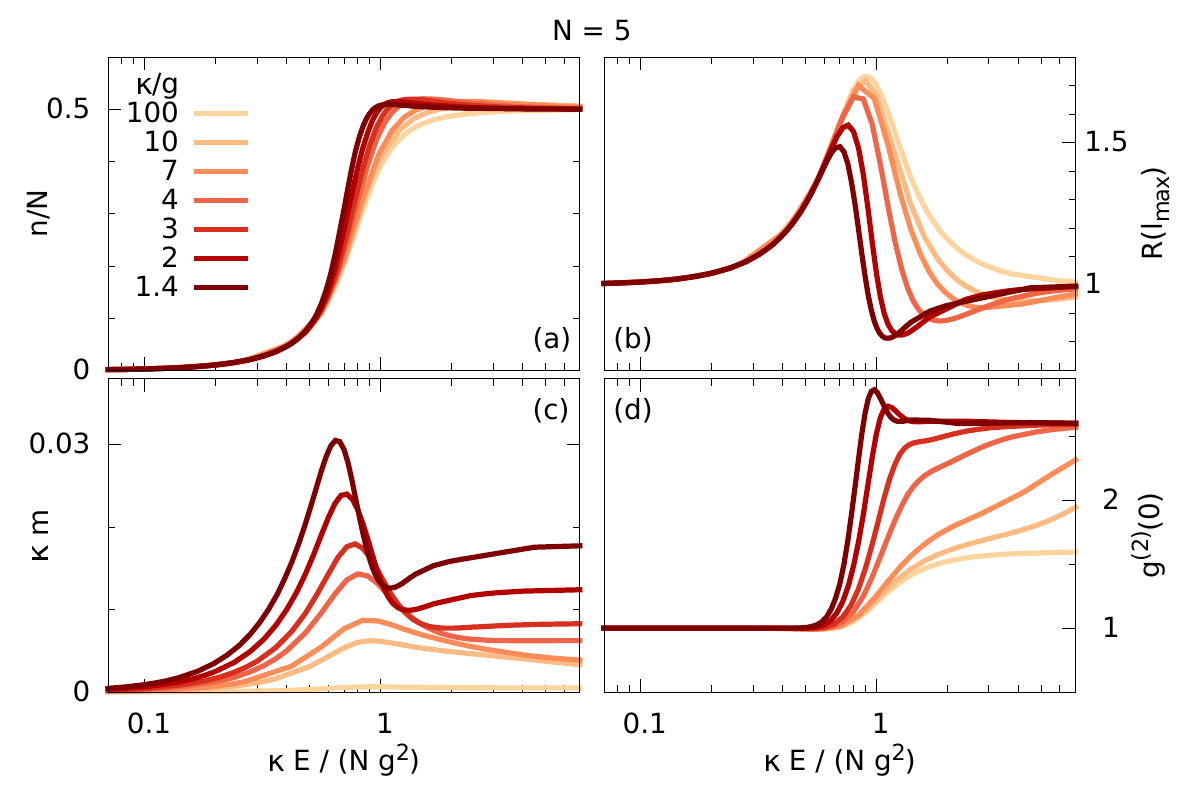}
\caption{Leaving the bad cavity limit: Variation of the external pumping strength for different ratios $\kappa/g$: (a) the normalized TLS excitation number $n/N=\langle J_{11} \rangle/N$, (b) the relative superradiant subspace occupation $R(l_{max}=N/2)$, (c) the cavity output rate $\kappa m= \kappa \langle b^\dagger b \rangle$ and (d) the photonic second order correlation function $g^{(2)}(0)$: Drastic qualitative change for $\kappa/g$ approaching unity.}
\label{fig.kappascale}
\end{figure}

\section{Results and Discussion}
We solve equation \eqref{eq.qme} with our computer library PsiQuaSP \cite{psiquasp,github} for master equations with reduced, polynomial scaling (see \ref{method} for a short introduction and Ref. \cite{Gegg:NJP:16} for more details). We use eigensolvers and time integration from PETSc and SLEPc \cite{petsc-web-page,petsc-user-ref,petsc-efficient,Hernandez:TMS:05}. \\
We use $\gamma = 1.0~\mbox{ns}^{-1}$ and $g = 3.3~\mbox{meV}$ throughout this work. Please note that ultra-strong coupling effects are not present in the investigated parameter range. There are two types of dephasing/individualization processes: spontaneous decay and pure dephasing. We first investigate the spontaneous decay and investigate the effects of pure dephasing later. Including small pure dephasing preserves all effects (see Section \ref{sec.robu} for a discussion).

\subsection{Nature of the phase transition}
\begin{figure}[t]
\centering
\includegraphics[scale=0.8]{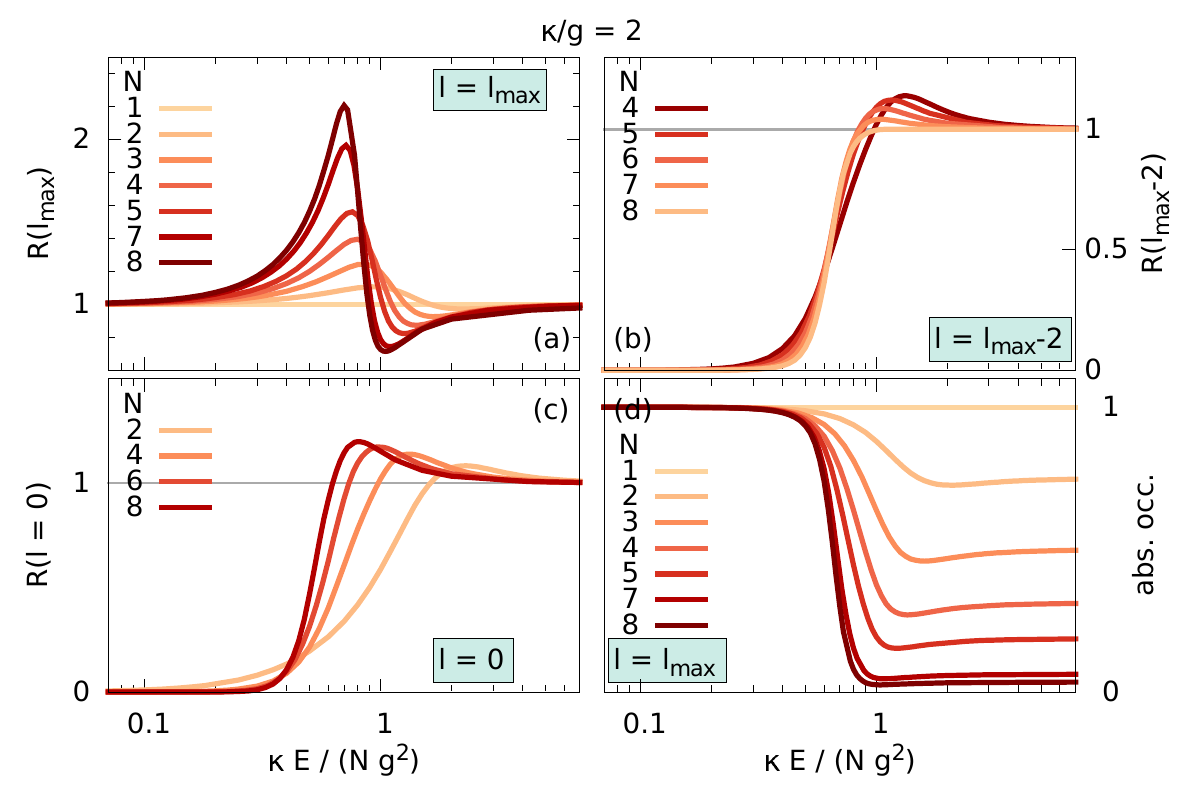}
\caption{Increasing the system size: Relative Dicke subspace occupation for varying $N$: (a) the superradiant subspace $l=N/2$, (b) $l=N/2 - 2$, (c) $l = 0$. These states have no interactions due to the Hamiltonian. They only couple to states with $l > 0$ through decay and dephasing. (d) Absolute occupation in the superradiant subspace: Approaching zero above the phase transition for $N\rightarrow\infty$, even without correlations.}
\label{fig.nscaledicke}
\end{figure}

\begin{figure}[t]
\centering
\includegraphics[scale=0.8]{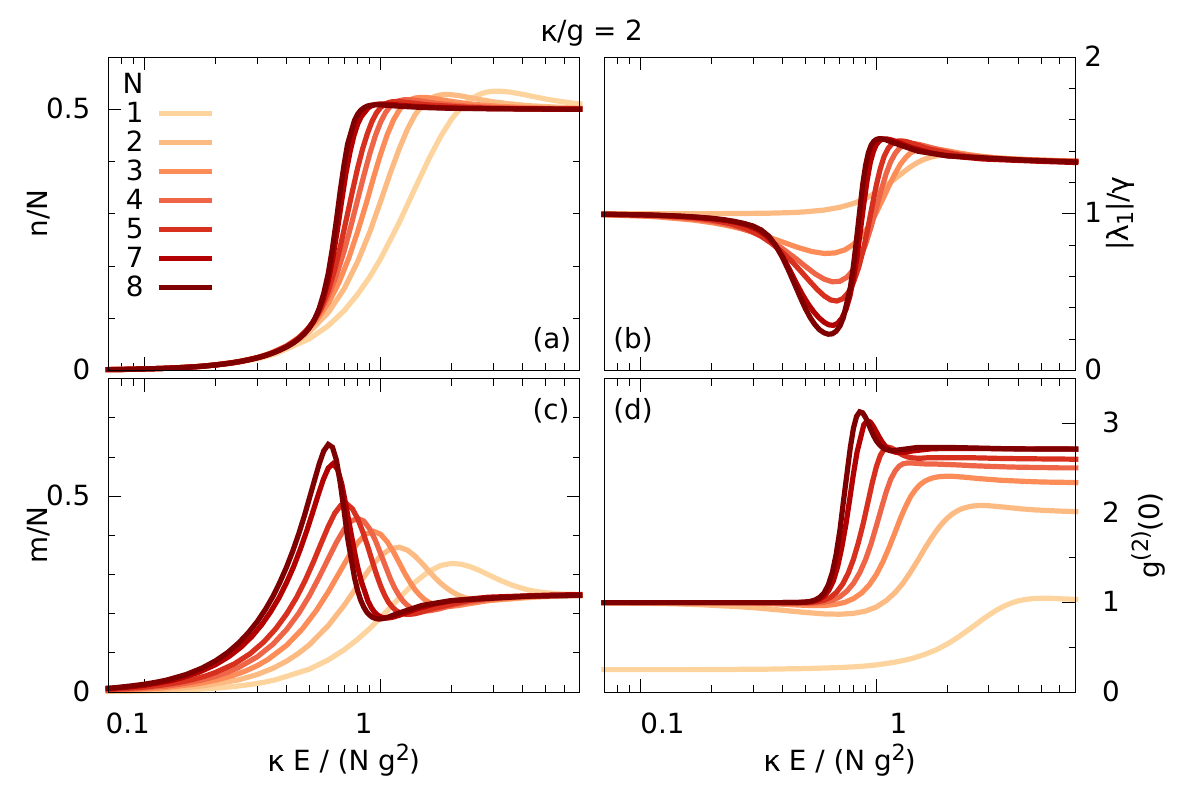}
\caption{Experimental signatures for varying $N$: (a) the normalized TLS excitation number $n/N=\langle J_{11} \rangle/N$, (b) the renormalized Liouvillian gap $|\lambda_1|/\gamma$, (c) the rescaled intracavity photon number $m/N=\langle b^\dagger b \rangle/N$ and (d) the second order correlation $g^{(2)}(0)$.}
\label{fig.nscaleobservables}
\end{figure}

In the steady state the most basic feature of the nonequilibrium phase transition is the change from the ground state to a half excited TLS state with increasing external driving field (Fig. \ref{fig.kappascale} (a)). Increasing the cavity quality (decreasing the ratio between cavity decay rate and TLS-cavity coupling strength $\kappa/g$) makes the transition sharper but the overall effect does not change much. Contrary a drastic change is seen in the behavior of the collectivity measure for the superradiant subspace $R(l_{max}=N/2)$, Fig. \ref{fig.kappascale} (b). While in the bad cavity limit the superradiant subspace population is always increased by collective effects ($R(l_{max}) \geq 1$), we observe an increased suppression ($R(l_{max}) < 1$) of the superradiant subspace for increasing cavity lifetime/quality. This is accompanied by a drastic increase of coherent cavity photons below and a pronounced bunching at moderate photon numbers above the phase transition (Fig. \ref{fig.kappascale} (c) and (d)). The maximum in the second order photon correlation function indicates the transition point from increased to suppressed superradiant subspace occupation. Please note that the cavity decay does not lead to an effective dephasing/individualization contribution for the TLS, thus the population of subradiant states through different cavity lifetimes is a highly nontrivial effect.\\
Above the phase transition collectivity favors the most subradiant subspace $l_{min}$: The dependence of $R(l_{max})$ on the number of TLS $N$, Fig. \ref{fig.nscaledicke} (a), shows a growing collective change in population of the superradiant subspace for increasing $N$. In Fig. \ref{fig.nscaledicke} (b) the ratio $R(l_{max}-2)$ is plotted -- it switches from collective suppression below to collective increase above the transition (this subspace only exists for $N\geq 4$). However the collective increase in population decreases for increasing $N$. For $N=4,5$ there are three different $l$ subspaces: $l_{max}$, $l_{max}-1$ and $l_{max}-2$. Thus for $N=4,5$ the subspace $l_{max}-2$ corresponds to the most subradiant subspace i.e. $N=4$: $l_{min} = 4/2- 2 = 0$ and $N=5$: $l_{min} = 1/2$. In these two cases the collective increase in population is strongest. For larger $N$ subspaces with smaller $l$ exist, e.g. $N=6: l_{min} = l_{max}-3$. Looking at $R(0)$ (only defined for even $N$, always corresponds to the most subradiant subspace), Fig. \ref{fig.nscaledicke} (c), we see that the increase due to collective effects increases with $N$. Hence the collective increase is always most pronounced in the most subradiant subspace ($l_{min}$) above the phase transition. 
Remarkably, below the phase transition the subradiant subspaces are completely suppressed, c.f. Figs. \ref{fig.nscaledicke} (b), (c).\\
The total occupation in the superradiant subspace goes to zero above the phase transition for $N \rightarrow \infty$, Fig. \ref{fig.nscaledicke} (d). Naively we could associate this with subradiance. However for $E \rightarrow \infty$ the TLS are in a completely incoherent, equipartitioned state \cite{Richter:JCP:07} and the superradiant subspace is only depopulated since this subspace becomes very small compared to the full Hilbert (Liouville) space for large $N$. This is clearly not a collective effect. 
This illustrates that (in the steady state) it is impossible to distinguish between collective and individual behavior by using Dicke state occupations alone.\\
However by looking at both the absolute and relative populations we conclude that in the good cavity and large $N$ limit the system changes from a predominantly superradiant to a predominantly subradiant state at the phase transition. This constitutes the main result of this work.\\
In Fig. \ref{fig.nscaleobservables} the scaling of experimentally more accessible quantities with the number of individual TLS $N$ is presented: The normalized TLS excitation develops a kink for increasing $N$, indicating a second-order transition, Fig. \ref{fig.nscaleobservables} (a). The smallest magnitude nonzero eigenvalue $\lambda_1$ of the Liouville operator $\mathcal{L}$ (c.f. equation \eqref{eq.qme}), which corresponds to the slowest time scale in the system to reach steady state, decreases around the phase transition for increasing $N$, Fig. \ref{fig.nscaleobservables} (b). It might even vanish for $N\rightarrow\infty$, creating a second steady state. This could be measured for instance in a hysteresis cycle typical for optical bistability experiments \cite{Rodriguez:arxiv:16,Carr:PhysRevLett:13,Gibbs:PhysRevLett:76}. The intracavity mean photon number shows the formation of a local minimum at the transition and an increase in the peak intensity, Fig. \ref{fig.nscaleobservables} (c). Also bunching ($g^{(2)}(0) > 1$) increases for increasing $N$, Fig. \ref{fig.nscaleobservables} (d). Overall the transition becomes sharper and more pronounced for increasing $N$ and decreasing $\kappa/g$, since these parameters increase the system size. This displays a typical property of phase transitions, which are well defined only in the thermodynamic limit (infinite system size) and blur for small system sizes \cite{Walls:SPTP:78,Rice:PhysRevA:94,Gross::01}.\\

\subsection{Robustness test and entanglement properties} 
\label{sec.robu}
\begin{figure}[t]
\centering
\includegraphics[scale=0.8]{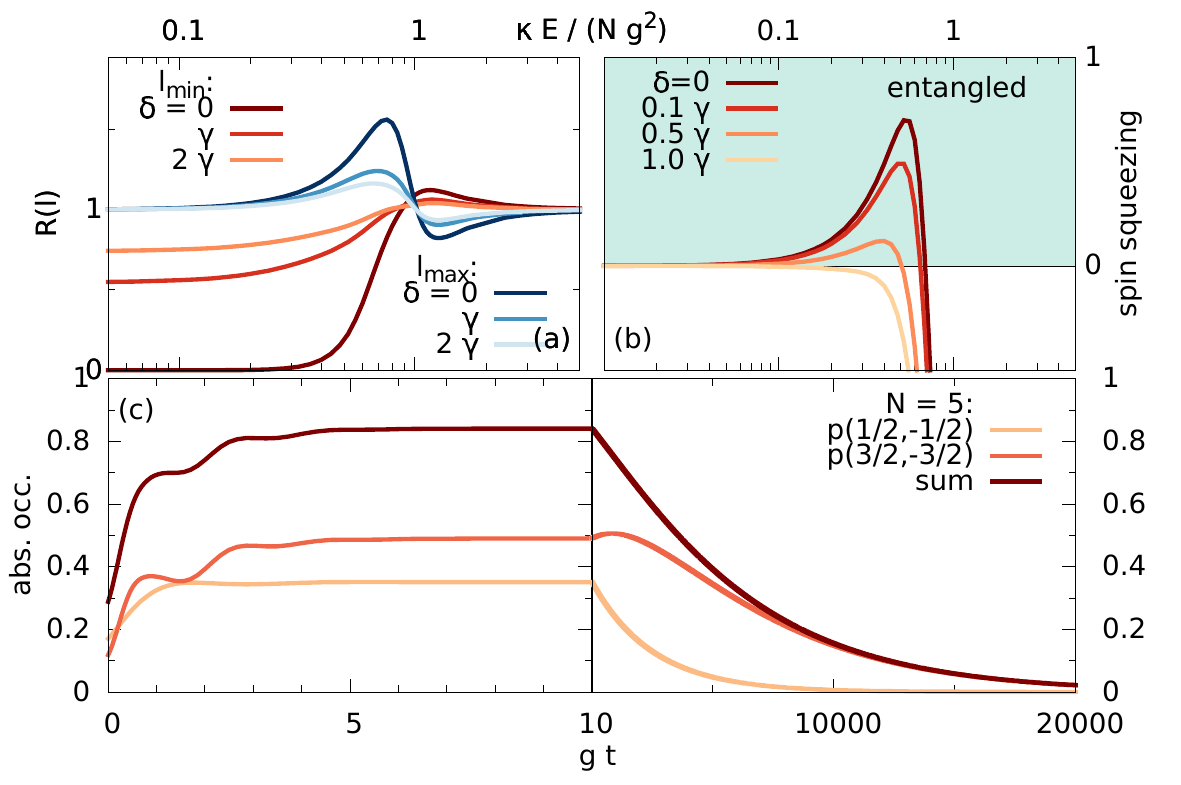}
\caption{Robustness, entanglement and dark state cascades: (a) The ratio $R(l)$ for $N = 5$ for $l=l_{min},l_{max}$ and varying $\delta$: The clear switching at the phase transition survives for $\delta \sim \gamma$. (b) Entanglement via spin squeezing inequalities: entanglement below the transition for $\delta < \gamma$. (c) Driving the system to the maximum subradiance point with subsequent relaxation to the ground state $N=5, \delta = 0$: A cascade of dark states is generated. Total dark state occupation close to unity.}
\label{fig.deltaspin}
\end{figure}

So far all results were presented without including pure dephasing. Now we investigate the robustness of the collective effects at the phase transition against pure dephasing: In Fig. \ref{fig.deltaspin} (a) we see that the collective behavior of the relative Dicke subspace population is reduced for increasing $\delta$. However the effect of clear distinction of superradiant state below and subradiant state above phase transition is preserved for $\delta \sim \gamma$. The general trend of total Dicke subspace occupation is not affected by pure dephasing, as in Fig. \ref{fig.nscaledicke} (d).\\
In the spin preserving setup the TLS are entangled via spin squeezing below the phase transition \cite{Gonzales:PhysRevLett:13}. Spin squeezing is a concept originating from quantum metrology, where it was developed around the idea that squeezed atomic coherent states could be used for measurement precision below the shot noise limit, but also has attracted a lot of attention as an entanglement witness \cite{Wineland:PhysRevA:92,Wineland:PhysRevA:94,Kitagawa:PhysRevA:93,Dylewsky:PhysRevA:16}. Here we employ the spin squeezing inequalities introduced by T\'oth \emph{et al.} that are explicitly derived as an entanglement witness for many two- (and multi-) level system setups \cite{Toth:PhysRevLett:07,Toth:PhysRevA:09}. The spin preserving case does not contain any subradiant states/effects and cannot model the effects of pure dephasing. The spin preserving and nonpreserving scenarios are two limits of the same physical system \cite{Agarwal:STMP:12,Carmichael::02}. Thus an investigation of entanglement in our setup and its preservation under dephasing is desirable: We find that the spin squeezing inequalities (SSI) by T\'oth \emph{et al.} detect entanglement below the phase transition for $\delta < \gamma$, see Fig. \ref{fig.deltaspin} (b) (see \ref{ssi} for the SSI and a definition of the quantity in Fig. \ref{fig.deltaspin} (b)). Hence the entanglement detected in the spin preserving setup is still present for spin nonpreserving setups and even for moderate pure dephasing times.

\subsection{Dark state cascades}
Super- and subradiance are concepts related to time evolution and so far we have only discussed the steady state: Now, we drive the system to the steady state with maximum $R(l_{min})$ (see Fig. \ref{fig.nscaledicke} (b) and (c)) and then, afterwards, we switch off the driving field. The system relaxes into the ground state and we observe that a cascade of dark states is generated, Fig. \ref{fig.deltaspin} (c): $p(1/2,-1/2)$ and $p(3/2,-3/2)$ are the populations in the lowest states of the smallest $l=l_{min}$ and intermediate $l_{max}>l>l_{min}$ subspace, c.f. Fig. \ref{fig.systemsketch} (b). Both states are dark. They are populated on time scales of the inverse TLS-photon coupling constant $g^{-1}$, because the higher energy, bright states of the associated $l$ subspaces decay via the emission of cavity photons. The cavity photons subsequently leave the cavity through the cavity decay. After the initial fast population of the $|l,-l\rangle$ states due to the TLS cavity interaction the dynamics are governed by spontaneous emission. The overall dark state population subsequently decays on the slower time scale $\gamma^{-1} = 5000 g^{-1}$ towards the ground state of the TLS ($|5/2,-5/2\rangle$). The decay follows the Dicke state cascade $p(1/2,-1/2) \rightarrow p(3/2,-3/2) \rightarrow p(5/2,-5/2)$. In general for different $N$: All $m > -l$ states relax to the $m=-l$ states on time scales of the inverse TLS-photon coupling constant $g^{-1}$ which is orders of magnitude faster than the decay time $\gamma^{-1}$. Subsequently the dark states $|l,-l\rangle$ relax in a cascade to the lower energy, dark states $|l+1,-l-1\rangle$ with minimal $l$ on time scales of $\gamma^{-1}$ towards the ground state $|l_{max},-l_{max}\rangle$, Fig. \ref{fig.systemsketch} (b). Please note that the overall occupation in subradiant dark states reaches values close to unity. Increasing the number of TLS also increases the dark state occupation during ground state relaxation. Subradiant correlations are clearly dominant here, since without these correlations the excitation in the TLS would still decay via the TLS cavity interaction Hamiltonian. This could be exploited for a controlled generation of subradiant states.

\section{Conclusion}
Experimental systems for observing the effects presented in this paper have to meet certain requirements: the pure dephasing of the TLS coherences should be small compared to the decay rate, i.e. $\delta \sim \gamma$. This can be realized with e.g. Rydberg ensembles \cite{Guerin:PhysRevLett:16,Pritchard:PRL:10,Saffman:RevModPhys:10} or with NV centers \cite{Tamarat:PhysRevLett:06} and quantum dots \cite{Borri:PhysRevLett:01} at low temperatures. Also a small inhomogeneous broadening is required, since it would likely blur the presented effect. For quantum dots this is more challenging than for NV centers and Rydberg ensembles. Generally, the decay rate $\gamma$ is not a crucial parameter but the ratio between decay and pure dephasing. If pure dephasing is too large the steady state effects are blurred, in the ground state relaxation subradiant state occupation is decreased and coherence times are shorter. However the dark state cascade effect is stable even against larger pure dephasing $\delta > \gamma$.\\
The parameters used in this study are realistic for NV centers, quantum dots and Rydberg atoms and the behavior is stable over a wide parameter range.\\
In summary we have shown that the nonequilibrium phase transition of cooperative resonance fluorescence changes drastically when leaving the bad cavity limit: Subradiant Dicke states are amplified and clear experimental signatures of this effect emerge. Letting the system relax into the ground state generates a dark state cascade that can be utilized to store quantum information.

\section{Acknowledgements}
We thank Nicolas Naumann for useful discussions, and gratefully acknowledge support from the Deutsche Forschungsgemeinschaft (DFG) through SFB 951 (M.G., M.R., A.K) and through the School of Nanophotonics of the SFB 787 (M.G.) and BR 1528/8-2 (A.C., A.K.).

\appendix
\section{Details to the permutation symmetric method}
\label{method}

The permutation symmetry of the master equation equation \eqref{eq.qme} confines the dynamics of the density matrix onto the subspace of symmetrized Liouville space states \cite{Hartmann:QIC:16,Gegg:NJP:16,psiquasp,github}:
\begin{equation}
\hat{\mathcal{P}}[n_{11},n_{10},n_{01}] = \mathcal{S} ~ \sigma_{11}^{\otimes n_{11}} \sigma_{10}^{\otimes n_{10}} \sigma_{01}^{\otimes n_{01}} \sigma_{00}^{\otimes n_{00}},
\label{eq.defbasis}
\end{equation} 
with $n_{00} = N -n_{11}-n_{10}-n_{01}$. The symmetrization operator is defined as $\mathcal{S} = \sum_P \hat{P}$, where $\hat{P}$ is the permutation operator and the sum is over all possible permutations $P$ of two-level systems. This expression is not normalized since the method is numerically more stable without normalization \cite{Gegg:NJP:16}. The density matrix can be expanded in the symmetric states using the Hilbert-Schmidt inner product
\begin{equation}
\mathcal{P}[n_{11},n_{10},n_{01}] = tr[\hat{\mathcal{P}}[n_{11},n_{10},n_{01}]\rho].
\end{equation}
Equations of motion can be derived from this expression by taking the time derivative and inserting the quantum master equation. In the PsiQuaSP library this is greatly facilitated by the use of a sketch representation for the symmetric basis states and the action of the Liouville space operators, there no derivation of equations of motion is required \cite{psiquasp,github}. The population in all states outside the symmetric Liouville subspace equation \eqref{eq.defbasis} is zero, if it is zero in the initial state. Compatible initial states are e.g. the ground state and the thermal equilibrium. The number of different symmetric basis states and thus the overall scaling of the method is $(N+1)(N+2)(N+3)/6 \sim N^3$. For $N=2$ we retrieve $10$ basis states. The $N=2$ states that occur in the Dicke state expansion are the classical occupation probabilities $\mathcal{P}[0,0,0] = \langle\sigma_{00}^1 \sigma_{00}^2\rangle$ (TLS ground state), $\mathcal{P}[1,0,0] = \langle \sigma_{11}^1 \sigma_{00}^2 + \sigma_{00}^1 \sigma_{11}^2\rangle$ (one TLS excited), $\mathcal{P}[2,0,0] = \langle \sigma_{11}^1 \sigma_{11}^2\rangle$ (both TLS excited) and the quantum correlation $\mathcal{P}[0,1,1] =\langle \sigma_{10}^1 \sigma_{01}^2 + \sigma_{01}^1 \sigma_{10}^2 \rangle$, with $\langle \dots \rangle = tr[\dots \rho]$. Exchanging the indices $1 \leftrightarrow 2$ leaves these states invariant -- they are permutation symmetric.\\
Continuing the example for $N=2$ the expectation values for the Dicke state projectors can be expanded in the symmetrized basis states: $p(1,-1)  = \mathcal{P}[0,0,0]$, $p(1,0) = 1/2\big(\mathcal{P}[1,0,0] + \mathcal{P}[0,1,1]\big)$, $p(1,1) = \mathcal{P}[2,0,0]$ and $p(0,0) = 1/2\big(\mathcal{P}[1,0,0] - \mathcal{P}[0,1,1]\big)$, using the trace condition of Ref. \cite{Xu:PhysRevA:13}.

\section{Spin sqeezing inequalities}
\label{ssi}

We employ the spin squeezing inequalities (SSI) introduced by T\'oth \emph{et al.} \cite{Toth:PhysRevLett:07,Toth:PhysRevA:09} as entanglement measure. T\'oth \emph{et al.} derived seven inequalities that are satisfied by any separable $N$-qubit state, hence the violation of any of these inequalities implies entanglement. Four of the seven inequalities detect entanglement in our setup, but the violation of two equations is equivalent: the coherent driving field introduces a time dependent phase factor caused by local unitary transformations which do not affect entanglement \cite{Vedral:PhysRevLett:97} but cause the violation of the SSI to oscillate back and forth between the two associated inequalities (between \eqref{eq.spinsqA1}, \eqref{eq.spinsqA2} and between \eqref{eq.spinsqB1}, \eqref{eq.spinsqB2}).
The four SSI that detect entanglement in our setup are
\begin{small}
 \begin{eqnarray}
\langle J_y^2\rangle + \langle J_z^2\rangle - \frac{N}{2} -(N-1)\big(\Delta J_x\big)^2 &\leq 0,\label{eq.spinsqA1}\\
\langle J_x^2\rangle + \langle J_z^2\rangle - \frac{N}{2} -(N-1)\big(\Delta J_y\big)^2 &\leq 0,\label{eq.spinsqA2}\\
\langle J_x^2\rangle + \frac{N(N-2)}{4} -(N-1)\left[ \big(\Delta J_y\big)^2 +\big(\Delta J_z\big)^2 \right]  &\leq 0,\label{eq.spinsqB1}\\
\langle J_y^2\rangle + \frac{N(N-2)}{4} -(N-1)\left[ \big(\Delta J_x\big)^2 +\big(\Delta J_z\big)^2 \right]  &\leq 0,\label{eq.spinsqB2}
\end{eqnarray}
\end{small}
where the variances are defined as $(\Delta A)^2 =\langle A^2 \rangle - \langle A \rangle^2 $.
In order to simplify the discussion we only show one SSI in our plot:
\begin{small}
 \begin{eqnarray}
\underbrace{\langle J_y^2\rangle + \langle J_z^2\rangle - \frac{N}{2} -(N-1)\big(\Delta J_x\big)^2}_{ =: A } &\leq 0,\label{eq.spinsqA}
\end{eqnarray}
\end{small}
hence $A$ is the quantity plotted in Fig. 5 (b).
Since strictly speaking the quantities $\langle J_y^2\rangle$ and $\langle J_x^2\rangle$ do not have a defined steady state, but oscillate with the phase factor mentioned above, we set $t=0$ and thus set the phase factor to unity throughout the plot in Fig. 5 (d). Since, as stated above, the local unitary transformations causing the oscillation do not affect the entanglement, this is a valid approach. In the following the local unitary transformation is explained:\\
On resonance the Hamiltonian of the system in a frame rotating at the external laser frequency $\omega_l$ reads
\begin{equation}
 H =  g (J_{10} b + J_{01} b^\dagger)  + E (J_{10} + J_{01}).
\end{equation}
The corresponding master equation for the setup considered in this work is
\begin{equation}
\partial_t \rho = \mathcal{L}\rho = \frac{i}{\hbar} [\rho,H] + \mathcal{D}_{de} + \mathcal{D}_{pd} + \mathcal{D}_{ph},
\end{equation}
where $\rho$ is the rotating frame density matrix. The transformation between normal frame and rotating frame is given by
\begin{equation}
 \rho_n = e^{-\frac{i}{\hbar} H_{\tiny{\mbox{rot}}}t} \rho e^{\frac{i}{\hbar} H_{\tiny{\mbox{rot}}}t},
\end{equation}
with the normal frame density matrix $\rho_n$ and the Hamiltonian
\begin{equation}
 H_{\tiny{\mbox{rot}}} =  \hbar \omega_l (b^\dagger b + J_{11}).
\end{equation}
The Hamiltonian acts locally on the density matrix, in the sense that each TLS experiences an individual unitary transformation, i.e.
\begin{equation}
 e^{J_{11}}= \prod_{i=1}^N e^{\sigma_{11}^i}.
\end{equation}
Such a transformation leaves the quantum correlations invariant \cite{Vedral:PhysRevLett:97}. Nonetheless some quantities arising in the SSI experience a time dependency through this transformation. In fact only the rotating frame density matrix has a stationary steady state, the normal frame density matrix $\rho_n$ exhibits an oscillating steady state, where diagonal entries are stationary and offdiagonal entries oscillate with a phase of multiples of $\omega_l$.\\
The quantities $\langle J_{x,y}^2 \rangle$ and $(\Delta J_{x,y})^2$ are explicitly time dependent in the normal frame. By adding Eqs. \eqref{eq.spinsqA1}, \eqref{eq.spinsqA2} and \eqref{eq.spinsqB1}, \eqref{eq.spinsqB2} respectively, one can derive time independent inequalities, which however do not detect entanglement in our setup.\\

\end{document}